\documentclass[sigchi,authorversion]{acmart}
\def\BibTeX{{\rm B\kern-.05em{\sc i\kern-.025em b}\kern-.08emT\kern-.1667em\lower.7ex\hbox{E}\kern-.125emX}}
    
\copyrightyear{2019}
\acmYear{2019}
\acmConference[MuC '19]{Mensch und Computer 2019}{September 8--11, 2019}{Hamburg, Germany}
\acmBooktitle{Mensch und Computer 2019 (MuC '19), September 8--11, 2019, Hamburg, Germany}
\acmPrice{15.00}
\acmDOI{10.1145/3340764.3344466}
\acmISBN{978-1-4503-7198-8/19/09}

\begin{document}

%
\title[Towards Collaborative Photorealistic VR Meeting Rooms]{Towards Collaborative Photorealistic VR Meeting Rooms}

%
\author{Alexander Sch\"afer}
\email{Alexander.Schaefer@dfki.uni-kl.de}
\affiliation{%
  \institution{TU Kaiserslautern}
}

\author{Gerd Reis}
\affiliation{%
  \institution{German Research Center for Artificial Intelligence}
}

\author{Didier Stricker}
\affiliation{%
  \institution{German Research Center for Artificial Intelligence}
  \institution{TU Kaiserslautern}
}

%
\renewcommand{\shortauthors}{Sch\"afer et al.}

%
\begin{abstract}
When designing 3D applications it is necessary to find a compromise between cost (e.g. money, time) and achievable realism of the virtual environment. Reusing existing assets has an impact on the uniqueness of the application and creating high quality 3D assets is very time consuming and expensive. We aim for a low cost, high quality and minimal time effort solution to create virtual environments. This paper's main contribution is a novel way of creating a virtual meeting application by utilizing augmented spherical images for photo realistic virtual environments.%
\end{abstract}

%
%

%

%
\begin{teaserfigure}
  \includegraphics[width=0.95\textwidth]{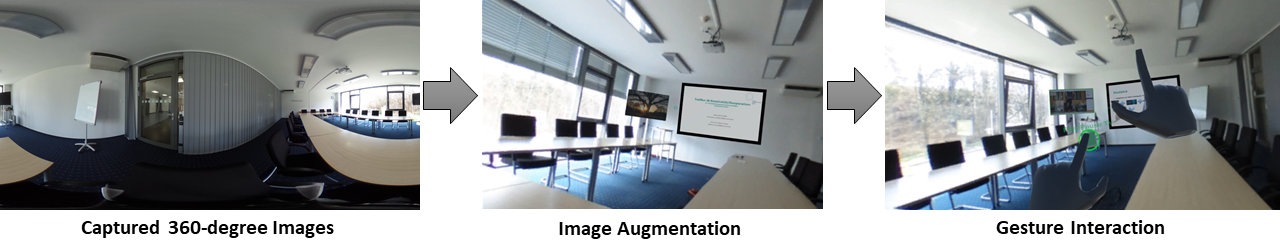}
  \caption{Our virtual reality meeting room using the proposed method.}
  \Description{Spherical images with virtual augmentation}
  \label{fig:teaser}
\end{teaserfigure}

%
\maketitle

\section{Introduction}
Meetings that require physical presence in other countries are often a time consuming overhead in addition to be costly. A great deal of money is spent on travel expenses while also reducing the working hours of employees by having them fly several hours from one end of the earth to the other.  In today's world, where everyone is connected via the internet and where it is possible to communicate very quickly with each other, such meetings should rarely be necessary. This is often not the case in reality. Back in 2008, IBM realized the potential of virtual conferences and saved approx. 320,000 \$ by having one of the biggest virtual conferences with over 200 participants according to Kantonen et al. \cite{Kantonen2010}. Andres et al. \cite{Andres2002} came to the conclusion that a physical face-to-face meeting is a far superior approach than having a video conference (e.g. Skype) in terms of productivity. The loss of social presence while having a phone or video call is an important point of their conclusion. Our work focuses on a virtual meeting scenario which provides tools for collaboration over long distances while maintaining a fully immersed experience. Our system allows an immersive experience to be created in mere minutes by using spherical images for virtual environments. Virtual objects are placed within those environments which can be viewed and interacted with by users in a virtually enriched experience (see Figure \ref{fig:teaser}). We use multiple spherical images from different view points to create an illusion which represents an authentic environment. Hand tracking is used as input and interaction method. During our work this approach has proven itself to be very successful in different setups while maintaining a high quality virtual environment with almost no effort.

The contribution of this work is as follows:
\begin{itemize}
    \item A novel VR application approach by utilizing multiple spherical images to provide a photo realistic enriched virtual world
    \item Utilizing hand gestures for interactive 3D elements in enriched virtual worlds in the context of virtual conferences
\end{itemize}

\begin{figure}
\includegraphics[width=0.85\columnwidth]{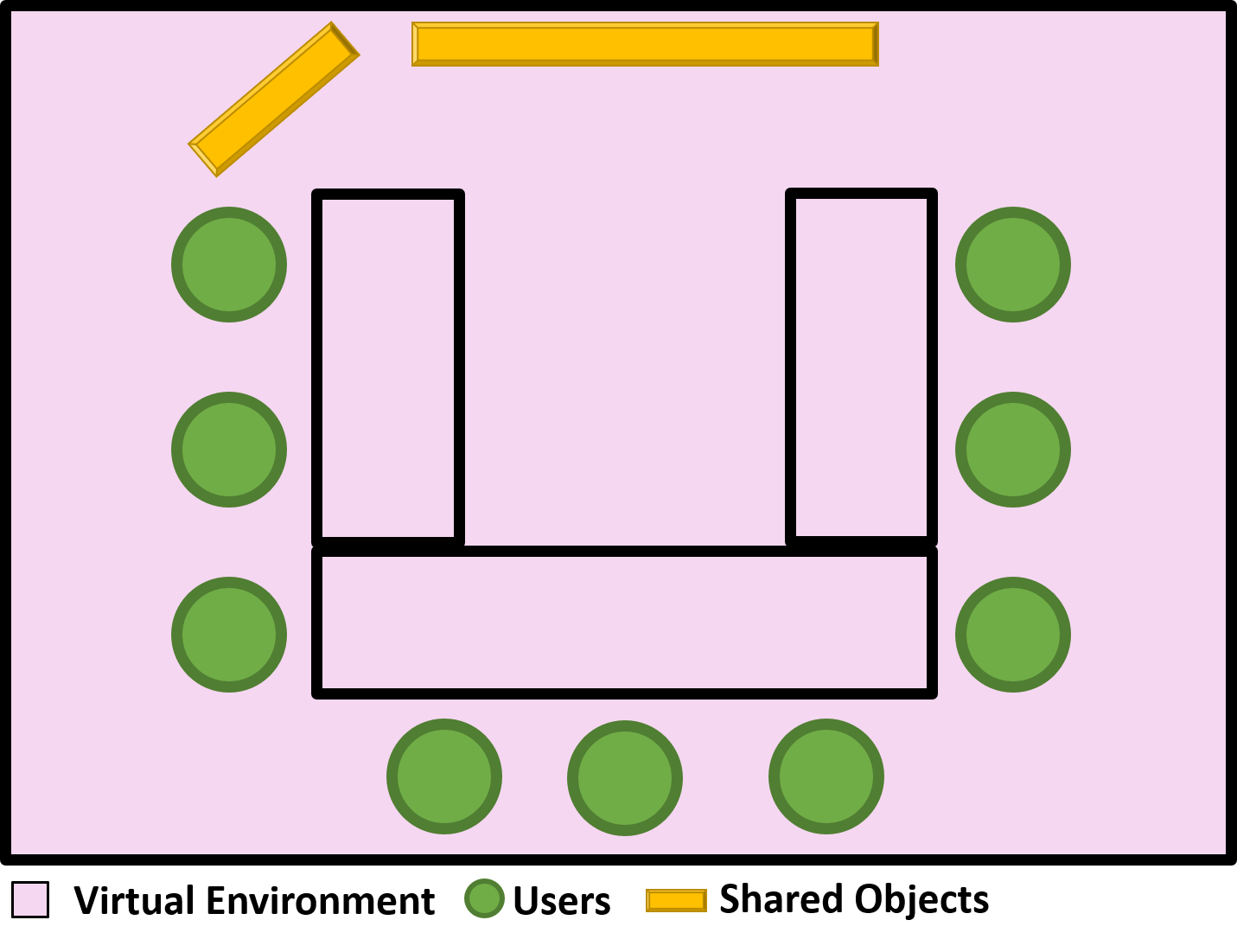}
\caption{Schematic overview of our approach. We separate three different roles in a virtual meeting.}
\label{fig:Illustration}
\end{figure}

\section{Related Work}

\begin{figure*}
\includegraphics[width=0.95\textwidth]{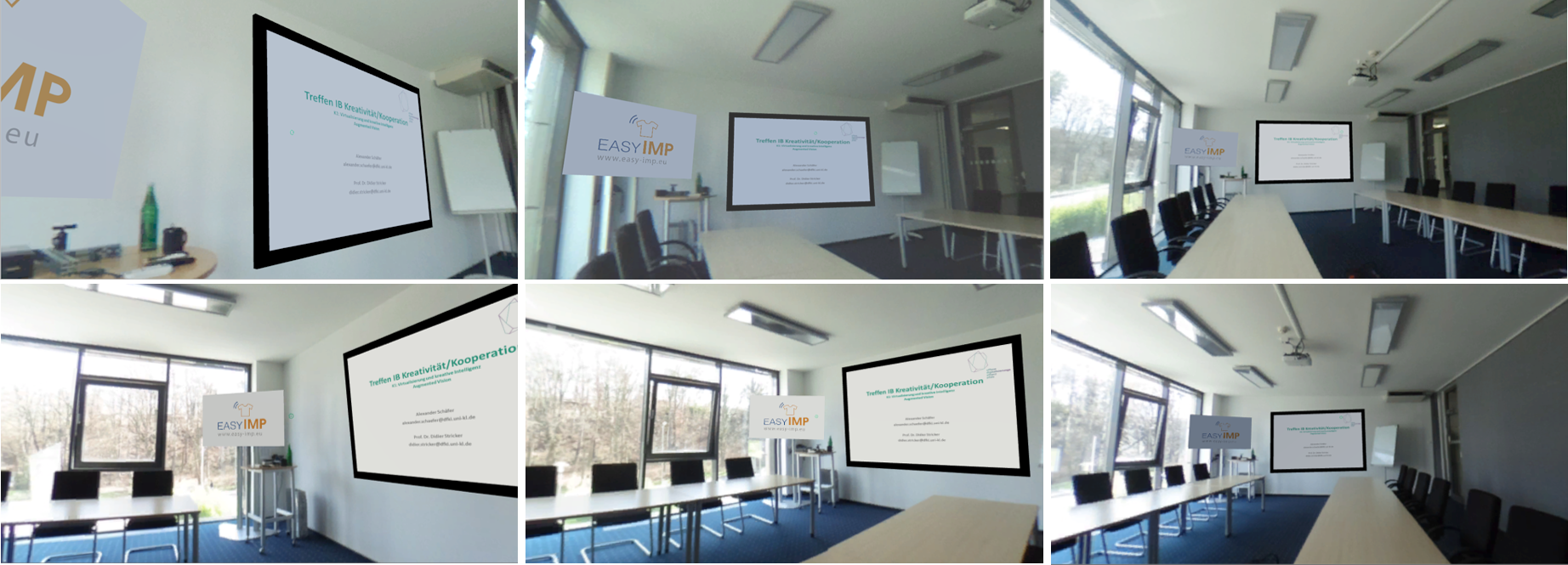}
\caption{Multiple spherical images share a virtual TV and projector.}
\label{fig:Viewpoints}
\end{figure*}

\subsection{Virtual Conferences}
A notably earlier work is from Kantonen et al. \cite{Kantonen2010} where they propose a system for teleconferencing by combining AR mechanics with the game environment Second Life, which was a very popular online social game. Jo et al. \cite{Jo2017} used 3D constructed environments and real environments (video background) in combination with different types of avatars. They showed that video based backgrounds are much more immersive and provide higher sense of co-presence than 3D replicated environments. A collaborative AR experience by using scene matching between two seperate rooms was also done by Jo et al. \cite{jo2015spacetime}. Their main component was a kinect sensor and a visual marker to match scenes and retarget avatar motions to fit in the physical place where each participant is located.

\subsection{Virtual Environments}
The presentation of the following works are mainly focused on the virtual environments used to create applications. Lovreglio et al. \cite{Lovreglio2018} used a complex 3D scene for prototyping a virtual earthquake simulation. While it allows a more fine grained control of the virtual environment, a modeled 3D scene lacks realism. Matsas et al. \cite{Matsas2018} created an application to analyze human robot collaboration in VR. They state that they used a virtual scene with at least 42 new 3D models created and image textures from real industrial workplaces for a more realistic environment while also adding several auxiliary parts and objects to the scene. While Matsas et al. \cite{Matsas2018} did not mention how much time it took to create this scene, creating this specific virtual environment must have been a large overhead. The work of Hilfert et. al \cite{hilfert2016low} tries to address this overhead issue while De Dinechin and Paljic et al. \cite{de2019automatic} mention 360-degree images as an easy-to-use and low-cost alternative to acquire a 3D scene. Virtual service journeys where evaluated by Boletsis et al. \cite{Boletsis2018}, where one evaluation method used 360-degree images from Google Street View for a prototyping phase. They mentioned that this approach was inexpensive in terms of man-hours and equipment costs compared to the other used prototype, a real world touring scenario. The Game Experience Questionnaire \cite{ijsselsteijn2013game} was used to evaluate a direct comparison between prototypes using the real environment and a virtual environment using 360-degree images of the visited places. They conclude that there was no big statistical difference in both prototypes which elevates the VR simulation prototype as a useful tool with significantly less expenses. A VR driving simulator for UI prototyping was done by Schroeter et al. \cite{Schroeter2018}. They used 180-degree videos and a high quality car model in an automated driving scenario for rapid prototyping of UI inside the car.

\subsection{Hand and Gesture Interaction Methods} 
There are several options available for gesture and hand based interaction. A Microsoft Kinect was used by Oikonomidis et al. \cite{oikonomidis2011efficient} for a markerless tracking of the hands. Hand tracking utilizing machine learning was done by Malik et al. \cite{malik2018structure}. A colored glove was used by Wang et al. \cite{wang2009real} to estimate hand pose from single RGB images. The leap motion controller (LMC) provides a tracking accuracy between 0.01mm and 0.5mm of the fingertip according to Smeragliuolo et al. \cite{Smeragliuolo2016}. We chose the LMC for our system since it is easy to integrate and  pose estimation processing is done on a dedicated device.

\section{Components for a virtual meeting room}
To create a virtual meeting room experience, we focused on three core aspects and their separate roles (illustrated in Figure \ref{fig:Illustration}): virtual environment, users, and shared interactive elements.
\newline\\
\textbf{Virtual Environments} are essential to immerse a user in virtual worlds. In our case it is desired to create a realistic experience that includes all important aspects of a common meeting room. A room with multiple seats, a big table and a projector is easily identified as a meeting room. The typical way of creating such a room is to create all necessary 3D assets and properly align them. Depending on the desired amount of realism this process can be very costly and time consuming. Another option is to take photos of a room and then recreate it with 3D assets. We found that approach to be too expensive and provide a solution which is independent of the complexity of a room and still provides very high realism. This is why we chose spherical images as a source to create photo realistic virtual worlds. The visual advantage compared to traditional artificially created 3D environments is enormous, they can capture the appearance of a room with every detail in a fracture of a second. Additionally it is possible to use HDR imaging techniques for even better visual impressions. Spherical images are one of our main components in creating a virtual meeting room.
\newline\\
\textbf{Users} need visual representation and the ability to interact with the virtual environment. Since the visualization method we use depends on real environments, we have aimed not to use any controllers and relied on gesture and hand based input methods, since it is a natural way of interaction in the real world. Users are represented by a virtual avatar in order to be perceived by others in the virtual environment. Seeing the hands and gestures of other users is a great way to enhance the overall collaboration experience \cite{piumsomboon2017exploring,sodhi2013bethere}. The hands are primarily used for interaction with virtual objects shared across all participants in the virtual meeting.
\newline\\
\textbf{Shared Interactive Elements} in meeting rooms are used to establish an information exchange between all persons present in a room. Those objects can be viewed, changed and interacted with by each participant. It is critical for a virtual meeting room experience to identify necessary components that enable a true meeting experience. In our case we restricted it to a projector and a TV to be those shared interactive elements, but this can be extended to any other element.

\section{Implementation}
\subsection{Overview}

Spherical images are used for the visualization of the virtual environment. We acquire spherical images of the resolution 5376 x 2688. Unity game engine was used for visualization and coupling of the hand tracking to our virtual simulation. The system supports all headsets supported by the OpenVR SDK (HTC Vive, Oculus Rift, ...).

We took several spherical images inside a meeting room, each centered at a spot where normally a participant would sit during a meeting. This enables the user to move to different positions in the room. Figure \ref{fig:Illustration} sketches our approach while Figure \ref{fig:Viewpoints} shows the result of the applied approach.

\begin{figure}
\includegraphics[width=0.95\columnwidth]{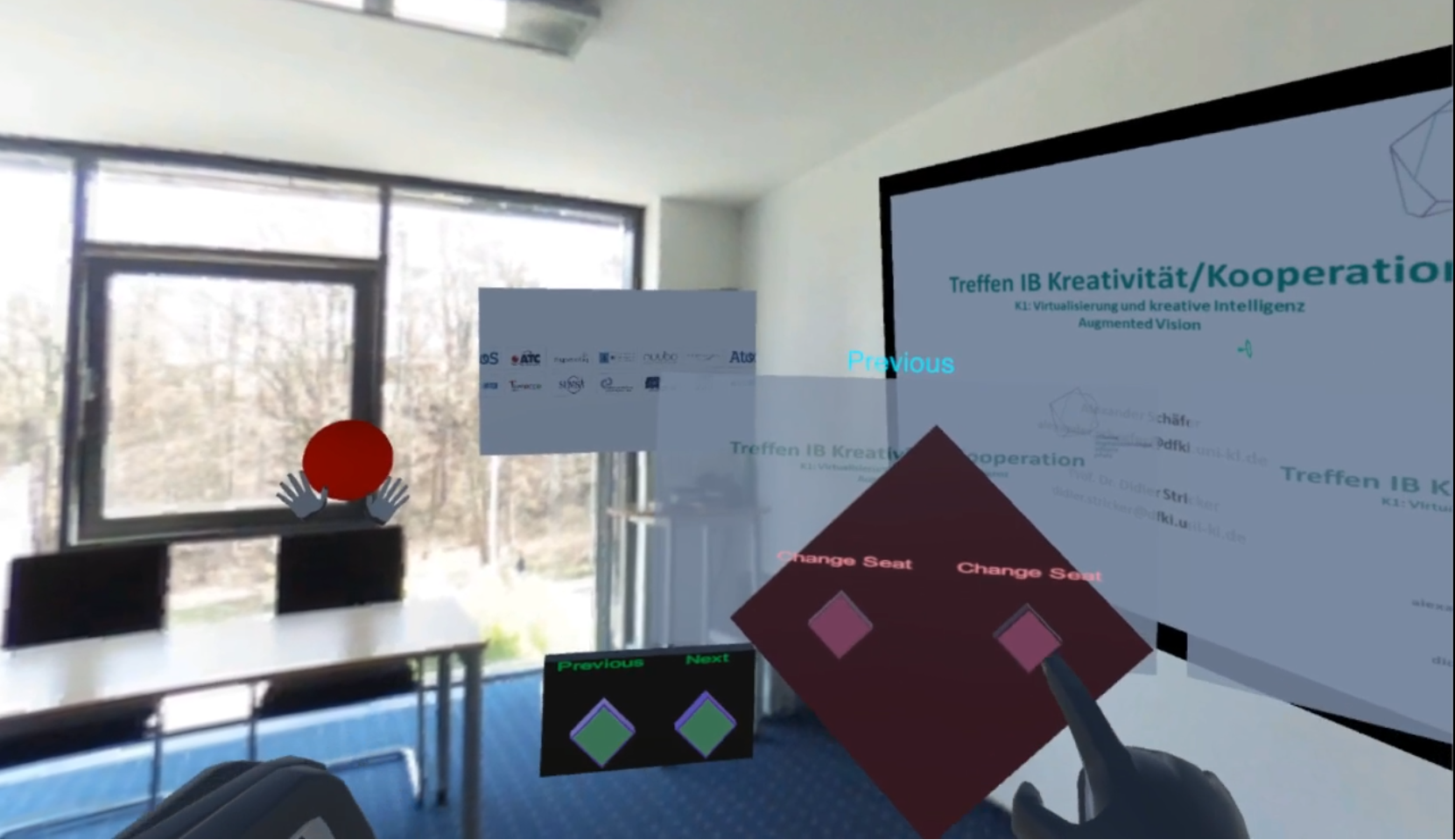}
\caption{Virtual objects can be placed by the user in the virtual environment to enable interaction.}
\label{fig:interaction}
\end{figure}

For collaboration, we implemented two possible tools inside our meeting room, a TV and a surface where a projector could project images on the wall. Those two tools can be interpreted as shared interactive elements to be seen by several users at the same time during a meeting. The identified objects share the same content for each participant but are placed at different positions inside the spherical viewer (see Figure \ref{fig:Viewpoints}). The current implementation uses manual placement of the shared elements. This will be extended to an autonomous process by using camera pose estimation methods. Multiple users can interact and observe the actions of each other in the meeting room. Currently, users see each other as a minimalistic avatar with hands attached to them (see Figure \ref{fig:interaction}). The fully rigged hands from the leap motion controller are streamed over network to enable users to see each others exact hand movements. For some interactions we simulated eye gaze by using a forward vector of the VR headset position in virtual space. This is used for certain gestures e.g. a swiping gesture towards the projector will change slides only when the projector is being looked at. We also use a menu which appears when the left palm is facing the VR headset. The user can grab objects from the menu and place them in the virtual world, where they expand and allow interaction (See Figure \ref{fig:interaction}). Interaction possibilities include: moving to a different seat, changing projector slides and preview for upcoming slides.

\subsection{Multi User Experience}
To implement a multi user experience, it is necessary to distinguish between the local and remote (other) users. We place the local user always at position (x,y,z) = (0,0,0) in the world coordinate system. A sphere which has the spherical image as texture applied to it is placed on that position to achieve a spherical viewer experience for the local user. In order to plausibly position the other users, it is important to know the camera pose of the other viewing points i.e. photos taken at other positions. Gava et al. \cite{gava2015generalized} provides an overview of the method we used for automated camera position detection. Without proper camera alignment the user positions will not be consistent. Therefore it is required to calibrate all user spheres into a common coordinate system. Each possible viewing position requires the following information: image texture, pose of shared interactive elements and pose of the other viewing points (user positions). Our network code synchronizes the position of players and the states of interactive elements as well as the fully rigged hands. We deemed it necessary to see both hands and full motion of each of the other users hands since it is an improvement for the overall user experience of the system.

\subsection{Creating a new VR Meeting Room}
Creating a virtual reality meeting room experience for a new room has few prerequirements:
\begin{itemize}
    \item Take images on desired seating positions
    \item Pose calibration of each seating position.
    \item Calibrate the pose of shared interactive elements
\end{itemize}
Taking images of the desired seating positions is straightforward but it has proven to be better if rotation stays the same. Also the height of the camera should be at eye level in seating position. Identifying the pose of each seating position can be done with an automated camera calibration framework. Identifying the pose of shared interactive elements has to be done manually at this time but can be extended to an automated process by utilizing camera pose estimation methods as well as sparse 3D reconstruction of the environment. Necessary components like gesture interaction, networking, visual representation of shared objects, spherical viewer, user avatars etc. can be reused easily.

\section{Conclusion and Future Work}

In this paper, we presented a method for fast and realistic creation of immersive virtual environments. To avoid creating complex 3D assets which are expensive, spherical images are used to create a photo realistic environment. The idea to use multiple spherical images showing the same scene but from different viewpoints while adding shared virtual objects between viewpoints is presented. This approach enables realistic and believable virtual environments which can be created in a short amount of time. Additionally, our approach allows us to create virtual meeting experiences in almost any meeting room by  taking multiple spherical images. Since we used the game engine Unity, it is easily possible to reuse all of the critical components like spherical image viewer, shared virtual objects, hand based gesture interaction and networking code. The presented system provides the opportunity for all basic interactions necessary to give a virtual presentation, including shared virtual objects (projector, TV) between viewpoints and hand/gesture interaction. This approach is easily extended for prototyping and proof-of-concept in other scenarios like novel car interfaces or training scenarios. In the current state it is possible for multiple users to collaborate together in the room but there is only a minimal visual representation of each user. The addition of more suitable avatars as well as sparse reconstruction to realize faithful occlusion of more complex avatars will round off the collaboration experience.
%

\begin{acks}
This work was partially funded by the Offene Digitalisierungsallianz Pfalz which is part of the Innovative Hochschule. Grant Number 03IHS075B.
\end{acks}

%
\bibliographystyle{ACM-Reference-Format}
\bibliography{sample-base}


\begin{thebibliography}{18}


\ifx \showCODEN    \undefined \def \showCODEN     #1{\unskip}     \fi
\ifx \showDOI      \undefined \def \showDOI       #1{#1}\fi
\ifx \showISBNx    \undefined \def \showISBNx     #1{\unskip}     \fi
\ifx \showISBNxiii \undefined \def \showISBNxiii  #1{\unskip}     \fi
\ifx \showISSN     \undefined \def \showISSN      #1{\unskip}     \fi
\ifx \showLCCN     \undefined \def \showLCCN      #1{\unskip}     \fi
\ifx \shownote     \undefined \def \shownote      #1{#1}          \fi
\ifx \showarticletitle \undefined \def \showarticletitle #1{#1}   \fi
\ifx \showURL      \undefined \def \showURL       {\relax}        \fi
\providecommand\bibfield[2]{#2}
\providecommand\bibinfo[2]{#2}
\providecommand\natexlab[1]{#1}
\providecommand\showeprint[2][]{arXiv:#2}

\bibitem[\protect\citeauthoryear{Andres}{Andres}{2002}]%
        {Andres2002}
\bibfield{author}{\bibinfo{person}{Hayward~P. Andres}.}
  \bibinfo{year}{2002}\natexlab{}.
\newblock \showarticletitle{{A comparison of face-to-face and virtual software
  development teams}}.
\newblock \bibinfo{journal}{\emph{Team Performance Management: An International
  Journal}}  \bibinfo{volume}{8} (\bibinfo{year}{2002}),
  \bibinfo{pages}{39--48}.
\newblock
\showISSN{13527592}
\urldef\tempurl%
\url{https://doi.org/10.1108/13527590210425077}
\showDOI{\tempurl}


\bibitem[\protect\citeauthoryear{Boletsis}{Boletsis}{2018}]%
        {Boletsis2018}
\bibfield{author}{\bibinfo{person}{Costas Boletsis}.}
  \bibinfo{year}{2018}\natexlab{}.
\newblock \showarticletitle{{Virtual Reality for Prototyping Service
  Journeys}}.
\newblock \bibinfo{journal}{\emph{Multimodal Technologies and Interaction}}
  \bibinfo{volume}{2}, \bibinfo{number}{2} (\bibinfo{year}{2018}),
  \bibinfo{pages}{14}.
\newblock
\urldef\tempurl%
\url{https://doi.org/10.3390/mti2020014}
\showDOI{\tempurl}


\bibitem[\protect\citeauthoryear{de~Dinechin and Paljic}{de~Dinechin and
  Paljic}{2019}]%
        {de2019automatic}
\bibfield{author}{\bibinfo{person}{Gregoire~Dupont de Dinechin} {and}
  \bibinfo{person}{Alexis Paljic}.} \bibinfo{year}{2019}\natexlab{}.
\newblock \showarticletitle{Automatic Generation of Interactive 3D Characters
  and Scenes for Virtual Reality from a Single-Viewpoint 360-Degree Video}.
\newblock  (\bibinfo{year}{2019}).
\newblock
\urldef\tempurl%
\url{https://doi.org/10.1109/VR.2019.8797969}
\showDOI{\tempurl}


\bibitem[\protect\citeauthoryear{Gava and Stricker}{Gava and Stricker}{2015}]%
        {gava2015generalized}
\bibfield{author}{\bibinfo{person}{Christiano~Couto Gava} {and}
  \bibinfo{person}{Didier Stricker}.} \bibinfo{year}{2015}\natexlab{}.
\newblock \showarticletitle{A generalized structure from motion framework for
  central projection cameras}. In \bibinfo{booktitle}{\emph{International Joint
  Conference on Computer Vision, Imaging and Computer Graphics}}. Springer,
  \bibinfo{pages}{256--273}.
\newblock


\bibitem[\protect\citeauthoryear{Hilfert and K{\"o}nig}{Hilfert and
  K{\"o}nig}{2016}]%
        {hilfert2016low}
\bibfield{author}{\bibinfo{person}{Thomas Hilfert} {and}
  \bibinfo{person}{Markus K{\"o}nig}.} \bibinfo{year}{2016}\natexlab{}.
\newblock \showarticletitle{Low-cost virtual reality environment for
  engineering and construction}.
\newblock \bibinfo{journal}{\emph{Visualization in Engineering}}
  \bibinfo{volume}{4}, \bibinfo{number}{1} (\bibinfo{year}{2016}),
  \bibinfo{pages}{2}.
\newblock


\bibitem[\protect\citeauthoryear{IJsselsteijn, De~Kort, and Poels}{IJsselsteijn
  et~al\mbox{.}}{2013}]%
        {ijsselsteijn2013game}
\bibfield{author}{\bibinfo{person}{WA IJsselsteijn}, \bibinfo{person}{YAW
  De~Kort}, {and} \bibinfo{person}{Karolien Poels}.}
  \bibinfo{year}{2013}\natexlab{}.
\newblock \showarticletitle{The game experience questionnaire}.
\newblock \bibinfo{journal}{\emph{Eindhoven: Technische Universiteit
  Eindhoven}} (\bibinfo{year}{2013}).
\newblock


\bibitem[\protect\citeauthoryear{Jo, Kim, and Kim}{Jo et~al\mbox{.}}{2015}]%
        {jo2015spacetime}
\bibfield{author}{\bibinfo{person}{Dongsik Jo}, \bibinfo{person}{Ki-Hong Kim},
  {and} \bibinfo{person}{Gerard~Jounghyun Kim}.}
  \bibinfo{year}{2015}\natexlab{}.
\newblock \showarticletitle{SpaceTime: adaptive control of the teleported
  avatar for improved AR tele-conference experience}.
\newblock \bibinfo{journal}{\emph{Computer Animation and Virtual Worlds}}
  \bibinfo{volume}{26}, \bibinfo{number}{3-4} (\bibinfo{year}{2015}),
  \bibinfo{pages}{259--269}.
\newblock


\bibitem[\protect\citeauthoryear{Jo, Kim, and Kim}{Jo et~al\mbox{.}}{2017}]%
        {Jo2017}
\bibfield{author}{\bibinfo{person}{Dongsik Jo}, \bibinfo{person}{Ki-hong Kim},
  {and} \bibinfo{person}{Gerard~Jounghyun Kim}.}
  \bibinfo{year}{2017}\natexlab{}.
\newblock \showarticletitle{{Effects of Avatar and Background Types on Users'
  Co-presence and Trust for Mixed Reality-Based Teleconference Systems}}.
\newblock \bibinfo{journal}{\emph{Casa 2017}} (\bibinfo{year}{2017}),
  \bibinfo{pages}{27--36}.
\newblock


\bibitem[\protect\citeauthoryear{Kantonen, Woodward, and Katz}{Kantonen
  et~al\mbox{.}}{2010}]%
        {Kantonen2010}
\bibfield{author}{\bibinfo{person}{Tuomas Kantonen}, \bibinfo{person}{Charles
  Woodward}, {and} \bibinfo{person}{Neil Katz}.}
  \bibinfo{year}{2010}\natexlab{}.
\newblock \showarticletitle{{Mixed reality in virtual world teleconferencing}}.
\newblock \bibinfo{journal}{\emph{Proceedings - IEEE Virtual Reality}}
  \bibinfo{number}{October} (\bibinfo{year}{2010}), \bibinfo{pages}{179--182}.
\newblock
\showISBNx{9781424462582}
\urldef\tempurl%
\url{https://doi.org/10.1109/VR.2010.5444792}
\showDOI{\tempurl}


\bibitem[\protect\citeauthoryear{Lovreglio, Gonzalez, Feng, Amor, Spearpoint,
  Thomas, Trotter, and Sacks}{Lovreglio et~al\mbox{.}}{2018}]%
        {Lovreglio2018}
\bibfield{author}{\bibinfo{person}{Ruggiero Lovreglio},
  \bibinfo{person}{Vicente Gonzalez}, \bibinfo{person}{Zhenan Feng},
  \bibinfo{person}{Robert Amor}, \bibinfo{person}{Michael Spearpoint},
  \bibinfo{person}{Jared Thomas}, \bibinfo{person}{Margaret Trotter}, {and}
  \bibinfo{person}{Rafael Sacks}.} \bibinfo{year}{2018}\natexlab{}.
\newblock \showarticletitle{{Prototyping virtual reality serious games for
  building earthquake preparedness: The Auckland City Hospital case study}}.
\newblock \bibinfo{journal}{\emph{Advanced Engineering Informatics}}
  \bibinfo{volume}{38} (\bibinfo{year}{2018}), \bibinfo{pages}{670--682}.
\newblock
\showISSN{14740346}
\urldef\tempurl%
\url{https://doi.org/10.1016/j.aei.2018.08.018}
\showDOI{\tempurl}
\showeprint[arxiv]{1802.09119}


\bibitem[\protect\citeauthoryear{Malik, Elhayek, and Stricker}{Malik
  et~al\mbox{.}}{2018}]%
        {malik2018structure}
\bibfield{author}{\bibinfo{person}{Jameel Malik}, \bibinfo{person}{Ahmed
  Elhayek}, {and} \bibinfo{person}{Didier Stricker}.}
  \bibinfo{year}{2018}\natexlab{}.
\newblock \showarticletitle{Structure-Aware 3D Hand Pose Regression from a
  Single Depth Image}. In \bibinfo{booktitle}{\emph{International Conference on
  Virtual Reality and Augmented Reality}}. Springer, \bibinfo{pages}{3--17}.
\newblock


\bibitem[\protect\citeauthoryear{Matsas, Vosniakos, and Batras}{Matsas
  et~al\mbox{.}}{2018}]%
        {Matsas2018}
\bibfield{author}{\bibinfo{person}{Elias Matsas},
  \bibinfo{person}{George~Christopher Vosniakos}, {and}
  \bibinfo{person}{Dimitris Batras}.} \bibinfo{year}{2018}\natexlab{}.
\newblock \showarticletitle{{Prototyping proactive and adaptive techniques for
  human-robot collaboration in manufacturing using virtual reality}}.
\newblock \bibinfo{journal}{\emph{Robotics and Computer-Integrated
  Manufacturing}} \bibinfo{volume}{50}, \bibinfo{number}{October}
  (\bibinfo{year}{2018}), \bibinfo{pages}{168--180}.
\newblock
\showISSN{07365845}
\urldef\tempurl%
\url{https://doi.org/10.1016/j.rcim.2017.09.005}
\showDOI{\tempurl}


\bibitem[\protect\citeauthoryear{Oikonomidis, Kyriazis, and
  Argyros}{Oikonomidis et~al\mbox{.}}{2011}]%
        {oikonomidis2011efficient}
\bibfield{author}{\bibinfo{person}{Iason Oikonomidis},
  \bibinfo{person}{Nikolaos Kyriazis}, {and} \bibinfo{person}{Antonis~A
  Argyros}.} \bibinfo{year}{2011}\natexlab{}.
\newblock \showarticletitle{Efficient model-based 3D tracking of hand
  articulations using Kinect.}. In \bibinfo{booktitle}{\emph{BmVC}},
  Vol.~\bibinfo{volume}{1}. \bibinfo{pages}{3}.
\newblock


\bibitem[\protect\citeauthoryear{Piumsomboon, Day, Ens, Lee, Lee, and
  Billinghurst}{Piumsomboon et~al\mbox{.}}{2017}]%
        {piumsomboon2017exploring}
\bibfield{author}{\bibinfo{person}{Thammathip Piumsomboon},
  \bibinfo{person}{Arindam Day}, \bibinfo{person}{Barrett Ens},
  \bibinfo{person}{Youngho Lee}, \bibinfo{person}{Gun Lee}, {and}
  \bibinfo{person}{Mark Billinghurst}.} \bibinfo{year}{2017}\natexlab{}.
\newblock \showarticletitle{Exploring enhancements for remote mixed reality
  collaboration}. In \bibinfo{booktitle}{\emph{SIGGRAPH Asia 2017 Mobile
  Graphics \& Interactive Applications}}. ACM, \bibinfo{pages}{16}.
\newblock


\bibitem[\protect\citeauthoryear{Schroeter and Gerber}{Schroeter and
  Gerber}{2018}]%
        {Schroeter2018}
\bibfield{author}{\bibinfo{person}{Ronald Schroeter} {and}
  \bibinfo{person}{Michael~A. Gerber}.} \bibinfo{year}{2018}\natexlab{}.
\newblock \showarticletitle{{A Low-Cost VR-Based Automated Driving Simulator
  for Rapid Automotive UI Prototyping}}.
\newblock  (\bibinfo{year}{2018}), \bibinfo{pages}{248--251}.
\newblock
\showISBNx{9781450359474}
\urldef\tempurl%
\url{https://doi.org/10.1145/3239092.3267418}
\showDOI{\tempurl}


\bibitem[\protect\citeauthoryear{Smeragliuolo, Hill, Disla, and
  Putrino}{Smeragliuolo et~al\mbox{.}}{2016}]%
        {Smeragliuolo2016}
\bibfield{author}{\bibinfo{person}{Anna~H. Smeragliuolo},
  \bibinfo{person}{N.~Jeremy Hill}, \bibinfo{person}{Luis Disla}, {and}
  \bibinfo{person}{David Putrino}.} \bibinfo{year}{2016}\natexlab{}.
\newblock \showarticletitle{{Validation of the Leap Motion Controller using
  markered motion capture technology}}.
\newblock \bibinfo{journal}{\emph{Journal of Biomechanics}}
  \bibinfo{volume}{49}, \bibinfo{number}{9} (\bibinfo{year}{2016}),
  \bibinfo{pages}{1742--1750}.
\newblock
\showISSN{18732380}
\urldef\tempurl%
\url{https://doi.org/10.1016/j.jbiomech.2016.04.006}
\showDOI{\tempurl}


\bibitem[\protect\citeauthoryear{Sodhi, Jones, Forsyth, Bailey, and
  Maciocci}{Sodhi et~al\mbox{.}}{2013}]%
        {sodhi2013bethere}
\bibfield{author}{\bibinfo{person}{Rajinder~S Sodhi}, \bibinfo{person}{Brett~R
  Jones}, \bibinfo{person}{David Forsyth}, \bibinfo{person}{Brian~P Bailey},
  {and} \bibinfo{person}{Giuliano Maciocci}.} \bibinfo{year}{2013}\natexlab{}.
\newblock \showarticletitle{BeThere: 3D mobile collaboration with spatial
  input}. In \bibinfo{booktitle}{\emph{Proceedings of the SIGCHI Conference on
  Human Factors in Computing Systems}}. ACM, \bibinfo{pages}{179--188}.
\newblock


\bibitem[\protect\citeauthoryear{Wang and Popovi{\'c}}{Wang and
  Popovi{\'c}}{2009}]%
        {wang2009real}
\bibfield{author}{\bibinfo{person}{Robert~Y Wang} {and} \bibinfo{person}{Jovan
  Popovi{\'c}}.} \bibinfo{year}{2009}\natexlab{}.
\newblock \showarticletitle{Real-time hand-tracking with a color glove}.
\newblock \bibinfo{journal}{\emph{ACM transactions on graphics (TOG)}}
  \bibinfo{volume}{28}, \bibinfo{number}{3} (\bibinfo{year}{2009}),
  \bibinfo{pages}{63}.
\newblock


\end{thebibliography}

%
\appendix
\end{document}